\pdfoutput=1																					

\documentclass[final,3p,times,twocolumn]{elsarticle}
 \biboptions{comma,sort&compress}

\usepackage{here}
 \usepackage{graphicx}
  \usepackage{epsfig}

\newcommand{\no}{\noindent}
\newcommand{\myeq}[3]{\vspace{#2} \begin{equation} \hspace{#1} #3 \end{equation} \vspace{0cm}}
\newcommand{\vsp}[1]{\vspace*{#1}}
\newcommand{\hsp}[1]{\hspace*{#1}}
\newcommand{\Op}{\mathcal{O}}

\journal{Nuc. Phys. (Proc. Suppl.)}

\begin{document}

\begin{frontmatter}



\title{Constraints on 3-flavor QCD order parameters and light quark mass difference from $\eta\to3\pi$ decays}

 \author[label1]{Mari\'{a}n Koles\'{a}r\corref{cor1}}
  \address[label1]{Institute of Particle and Nuclear Physics, Faculty of Mathematics and Physics, Charles University in Prague, CZ-18000 Prague, Czech republic}
	\cortext[cor1]{Speaker}
	\author[label1]{Ji\v{r}\'i Novotn\'y}


\begin{abstract}

The $\eta$$\,\to\,$$3\pi$ decays are a valuable source of information on low energy QCD. Yet they were not used for an extraction of the chiral symmetry breaking order parameters until now.  We use a bayesian approach in the framework of resummed chiral perturbation theory to obtain constraints on the quark condensate and pseudoscalar decay constant in the chiral limit, as well as the mass difference of light quarks. We compare our results with recent $\chi$PT and lattice QCD fits and find some tension, as the $\eta$$\,\to\,$$3\pi$ data seem to prefer a larger ratio of the chiral order parameters. The results also seem to exclude a large value of the chiral decay constant, which was found by some recent works. 

\end{abstract}





\end{frontmatter}



Spontaneous breaking of chiral symmetry (SB$\chi$S) is a prominent feature of the QCD vacuum and thus its character has been under discussion for a long time \cite{Fuchs:1991cq,DescotesGenon:1999uh}. The principal order parameters are the quark condensate and the pseudoscalar decay constant in the chiral limit

\myeq{0cm}{0cm}{
	 \Sigma(N_f) = -\langle\,0\,|\,\bar{q}q \,|\,0\,\rangle\,|_{m_q\to 0}\,,}
\myeq{0cm}{-0.25cm}{
	 F(N_f) = F_P^a\,|_{m_q\to 0}\,,\quad   i p_{\mu}\, F_P^a\ =\ \langle\,0\,|\,A_{\mu}^a\,|\,P\,\rangle,}

\no where $N_f$ is the number of quark flavors $q$ considered light, $m_q$ collectively denotes their masses. $A_{\mu}^a$ are the QCD axial vector currents, $F_P^a$ the decay constants of the light pseudoscalar mesons $P$.

Chiral perturbation theory ($\chi$PT) \cite{Weinberg:1978kz,Gasser:1983yg,Gasser:1984gg} is constructed as a general low energy parameterization of QCD based on its symmetries and the discussed order parameters appear at the lowest order of the chiral expansion as low energy constants (LECs). Interactions of the light pseudoscalar meson octet, the pseudo-Goldstone bosons of the broken symmetry, directly depend on the pattern of SB$\chi$S and thus can provide information about the values of these observables.

A convenient reparameterization of the order parameters, relating them to physical quantities connected with pion two point Green functions, can be introduced \cite{DescotesGenon:1999uh}

\myeq{0.25cm}{0cm}{
	Z(N_f) = \frac{F(N_f)^2}{F_{\pi}^2},\quad
	X(N_f) = \frac{2\hat{m}\,\Sigma(N_f)}{F_{\pi}^2M_{\pi}^2},}

\no where $\hat{m}\,$=$\,(m_u\,$+$\,m_d)/2$. $X(N_f)$ and $Z(N_f)$ are limited to the range (0,\,1), $Z(N_f)$\,=\,0 would correspond to a restoration of chiral symmetry and $X(N_f)$\,=\,0 to a case with vanishing chiral condensate. Standard approach to chiral perturbation series tacitly assumes values of $X(N_f)$ and $Z(N_f)$ not much smaller than one, which means that the leading order terms should dominate the expansion.

Several recent results for the two and three flavor order parameters are listed in tables \ref{tab1} and \ref{tab2}, respectively. As can be seen, while the two flavor case is quite settled, the values of $X(2)$ and $Z(2)$ being not much smaller than one, the situation in the three flavor one is much less clear. Some analyses suggest a significant suppression of X(3) and/or Z(3) and thus a non-standard behavior of the spontaneously broken QCD vacuum.

\begin{table} \small \hsp{0.8cm}
\begin{tabular}{|c|c|c|c|}
	\hline \rule[-0.2cm]{0cm}{0.5cm} & $Z(2)$ & $X(2)$ \\
	\hline \rule[-0.2cm]{0cm}{0.5cm} $\pi\pi$ scattering \cite{DescotesGenon:2001tn} & 0.89$\pm$0.03 & 0.81$\pm$0.07 \\
	\rule[-0.2cm]{0cm}{0.5cm} lattice QCD \cite{Bernard:2012fw} & 0.86$\pm$0.01 & 0.89$\pm$0.01 \\
	\hline
\end{tabular}
	\caption{Chosen results for the two flavor order parameters.}
	\label{tab1}
\end{table}\normalsize

\begin{table} \small
\begin{tabular}{|c|c|c|c|}
	\hline \rule[-0.2cm]{0cm}{0.5cm} phenomenology & $Z(3)$ & $X(3)$ \\
	\hline \rule[-0.2cm]{0cm}{0.5cm} NNLO $\chi$PT (main fit) \cite{Bijnens:2014lea} & 0.59 & 0.63\\
	\rule[-0.2cm]{0cm}{0.5cm} NNLO $\chi$PT (free fit) \cite{Bijnens:2014lea} & 0.51 & 0.48 \\
	\rule[-0.2cm]{0cm}{0.5cm} NNLO $\chi$PT ("fit 10") \cite{Amoros:2001cp} & 0.89 & 0.66 \\
	\rule[-0.2cm]{0cm}{0.5cm} Re$\chi$PT $\pi\pi$+$\pi K$ \cite{DescotesGenon:2007ta} & $>$0.2 & $<$0.8 \\
	\hline \rule[-0.2cm]{0cm}{0.5cm} lattice QCD & $Z(3)$ & $X(3)$ \\
	\hline \rule[-0.2cm]{0cm}{0.5cm} RBC/UKQCD+Re$\chi$PT \cite{Bernard:2012ci} & 0.54$\pm$0.06 & 0.38$\pm$0.05\\
	\rule[-0.2cm]{0cm}{0.5cm} RBC/UKQCD+large $N_c$ \cite{Ecker:2013pba} & 0.91$\pm$0.08 & \\
	\rule[-0.2cm]{0cm}{0.5cm} MILC 09A \cite{Bazavov:2009fk} & 0.72$\pm$0.06 & 0.62$\pm$0.07 \\
	\hline
\end{tabular}
	\caption{Chosen results for the three flavor order parameters.}
	\label{tab2}
\end{table}\normalsize

Up, down and strange quark masses are other parameters governing the low energy QCD physics. A commonly used reparameterization can be used

\myeq{0cm}{0cm}{
	\hat{m} = \frac{m_u + m_d}{2},\quad
	r = \frac{m_s}{\hat{m}},\quad
	R = \frac{m_s-\hat{m}}{m_d-m_u}.}

\no The values for the light quark mass average and the strange to light quark mass ratio are well known from lattice QCD and QCD sum rules \cite{Aoki:2013ldr,Narison:2014vka}. On the other hand, the isospin breaking parameter $R$, directly related to the light quark mass difference, has not been reliably determined by these methods yet.

The $\eta$$\,\to\,$$3\pi$ isospin breaking decays have not been exploited for an extraction of the chiral order parameters so far, yet we argue there is valuable information to be had. The theory seems to converge slowly for the decays. One loop corrections were found to be very sizable \cite{Gasser:1984pr}, the result for the decay width of the charged channel was 160$\pm$50 eV, compared to the current algebra prediction of 66 eV. However, the experimental value is still much larger, the current PDG value is \cite{PDG:1900zz}

\myeq{1.75cm}{0cm}{\Gamma^+_\mathrm{exp} = 300 \pm 12 \ \mathrm{eV}.}

\no Only the two loop $\chi$PT calculation \cite{Bijnens:2007pr} has succeeded to obtain a reasonable result for the widths. The latest experimental value for the neutral decay width is \cite{PDG:1900zz}

\myeq{1.75cm}{0cm}{\Gamma^0_\mathrm{exp} = 428 \pm 17 \ \mathrm{eV}.}

As will be shown elsewhere \cite{Kolesar:prep}, we argue that the four point Green functions relevant for the $\eta$$\,\to\,$$3\pi$ amplitude (see (\ref{Green_f}) below) do not necessarily have large contributions beyond next-to-leading order and a reasonably small higher order remainder is not in contradiction with huge corrections to the decay widths. The widths do not seem to be sensitive to the details of the Dalitz plot distribution, but rather to the value of leading order parameters - the chiral decay constant, the chiral condensate and the difference of $u$ and $d$ quark masses, i.e. the magnitude of isospin breaking. Moreover, access to the values of these quantities is not screened by EM effects, it was shown that the electromagnetic corrections up to NLO are very small \cite{Baur:1995gc, Ditsche:2008cq}. This is our motivation for our effort to extract information about the character of the QCD vacuum from this decay.

The Dalitz plot distributions are experimentally well known as well \cite{KLOE:2008ht,KLOE:2010mj,WASA:2014aks}. However, as we will discuss in detail in \cite{Kolesar:prep}, we have not found the convergence of the theory in the case of the slopes reliable enough to include all the Dalitz plot parameters into the analysis. To stay on the conservative side, we used the lowest order parameter in the charged channel only \cite{KLOE:2008ht}

\myeq{1.75cm}{0cm}{a = -1.09 \pm 0.02.}

Our calculation closely follows the procedure outlined in \cite{Kolesar:2008jr}, results presented here are a significant update on our initial report \cite{Kolesar:2013ywa}. We use an alternative approach to chiral perturbation theory, dubbed resummed $\chi$PT (Re$\chi$PT) \cite{DescotesGenon:2003cg}, which was developed in order to accommodate the possibility of irregular convergence of the chiral expansion. The procedure can be very shortly summarized in the following way:

\vsp{-0.1cm}
\begin{itemize}	
	\item[-]	standard $\chi$PT Lagrangian and power counting \vsp{-0.2cm}
	\item[-]	only expansions related linearly to Green functions of the QCD currents trusted \vsp{-0.2cm}
	\item[-] 	explicitly to NLO, higher orders implicit\\ in remainders\vsp{-0.2cm}
	\item[-]	remainders retained, treated as sources of error \vsp{-0.2cm}
	\item[-]	manipulations in non-perturbative algebraic way \vsp{-0.1cm}
\end{itemize}

\no The hope for resummed $\chi$PT is that by carefully avoiding dangerous manipulations a better converging series can be obtained. The procedure also avoids the hard to control NLO a NNLO LECs by trading them for remainders with known chiral order.

We start by expressing the charged decay amplitude in terms of 4-point
Green functions $G_{ijkl}$, obtained from the generating functional of the QCD currents. The neutral decay amplitude can be straightforwardly obtained from the charged one. We compute at first order in isospin breaking, 
the amplitude then takes the form

\myeq{-0.7cm}{0cm}{\label{Green_f}
	F_\pi^3F_{\eta}A(s,t,u)
		= G_{+-83}-\varepsilon_{\pi}G_{+-33}+\varepsilon_{\eta}G_{+-88} + \Delta^{(6)}_{G_D},\ }
		
\no where $\Delta^{(6)}_{G_D}$ is the direct higher order remainder to the 4-point Green functions. 
The physical mixing angles to all chiral orders and first in isospin breaking
can be expressed in terms of quadratic mixing terms of the generating functional to NLO
and related indirect remainders

\myeq{-0.5cm}{0cm}{
	\varepsilon_{\pi,\eta} = -\frac{F_{0}^{2}}{F_{\pi^0,\eta}^{2}}
		\frac{(\mathcal{M}_{38}^{(4)}+\Delta_{M_{38}}^{(6)}) - 									
		M_{\eta,\pi^0}^{2}(Z_{38}^{(4)}+\Delta _{Z_{38}}^{(6)})}
		{M_\eta^2-M_{\pi^0}^2}.}
		
In accord with the method, $\Op(p^2)$ parameters appear inside loops, while
physical quantities in outer legs. Such a strictly derived amplitude has an
incorrect analytical structure due to the leading order masses in loops, cuts and poles being in unphysical positions. To account
for this, we exchange the LO masses in unitarity corrections and chiral logarithms for physical ones, as described in \cite{Kolesar:2008jr}.   

The next step is the treatment of the LECs. As discussed, the leading order
ones, as well as quark masses, are expressed in terms of convenient parameters $X$, $Z$, $r$ and $R$.
At next-to-leading order, the LECs $L_4$-$L_8$ are algebraically reparametrized in terms
of pseudoscalar masses, decay constants and the free parameters $X$, $Z$ and $r$ using chiral expansions of 
two point Green functions, similarly to \cite{DescotesGenon:2003cg}. Because expansions are formally not truncated, 
each generates an unknown higher order remainder.

We don't have a similar procedure ready for $L_1$-$L_3$ at this point, therefore we collect
a set of standard $\chi$PT fits \cite{Amoros:2001cp,Bijnens:2011tb,Bijnens:1994ie} and by taking their mean and spread, while 
ignoring the much smaller reported error bars, we obtain an estimate of their influence

\myeq{0.75cm}{0cm}{L_1^r(M_\rho) = (0.60\pm$$0.28) \cdot 10^{-3}}
\myeq{0.75cm}{-0.5cm}{L_2^r(M_\rho) = (0.88\pm$$0.34) \cdot 10^{-3}}
\myeq{0.75cm}{-0.5cm}{L_3^r(M_\rho) = (-2.97\pm$$0.47) \cdot 10^{-3}}

\no As will be shown in \cite{Kolesar:prep}, the results depend on these constants only very weakly.

The $O(p^6)$ and higher order LECs, notorious for their abundance, are implicit in
a relatively smaller number of higher order remainders. We have eight indirect remainders
- three generated by the expansions of the pseudoscalar masses, three by the decay constants and two by the mixing angles.
We expand the direct remainder to the 4-point Green functions around the center of the Dalitz plot $s_0=1/3(M_\eta^2$+$2M_{\pi^+}^2$+$M_{\pi^0}^2)$
		
\myeq{-0.5cm}{0cm}{
	\Delta^{(6)}_{G_D} = \Delta_A+\Delta_B(s-s_0)\,+}
\myeq{0.5cm}{-0.5cm}{+\,\Delta_C(s-s_0)^2+\Delta_D [(t-s_0)^2+(u-s_0)^2]}
	
\no and thus get four derived direct remainders, two NLO and two NNLO ones. As the experimental curvature of the Dalitz plot is very small \cite{KLOE:2008ht}, we argue that for our purpose of calculating the decay widths and the lowest order Dalitz slope $a$ the expansion to second order in the Mandelstam variables is sufficient. 

For the statistical analysis, we use an approach based on Bayes' theorem \cite{DescotesGenon:2003cg}

\myeq{0.5cm}{0cm}{\label{Bayes}
		P(X_i|\mathrm{data}) = \frac{\prod_k P(\mathrm{O_k}|X_i)P(X_i)}{\int \mathrm{d}X_i\,\prod_k 		
		P(\mathrm{O_k}|X_i)P(X_i)}\,,}

\no where $P(X_i|\mathrm{data})$ is the probability density of the parameters and remainders, denoted as $X_i$, having a specific value given the observed experimental data. $P(O_k|X_i)$ are the known probability densities of obtaining the observed values of the included observables $O_k$ in a set of independent experiments with uncertainties $\sigma_k$ under the assumption that the true values of $X_i$ are known

\myeq{-0.25cm}{0cm}{
		P(O_k|X_i) = \frac{1}{\sigma_k\sqrt{2\pi}}\,
		\mathrm{exp}\left[-\frac{(O_k-O^\mathrm{th}_k(X_i))^2}{\sigma_k}\right].}
		
\no Our observables, treated as independent, are the charged and neutral decay widths and the Dalitz slope $a$.

$P(X_i)$ are the prior probability distributions of $X_i$. We use them to implement the theoretical uncertainties connected with our parameters and remainders. In such a way we keep the theoretical assumptions explicit and under control. This also allows us to test various assumptions and formulate if-then statements as well as implement additional constraints (see below).

We assume the strange to light quark ratio $r$ to be known and use the lattice QCD average \cite{Aoki:2013ldr}

\myeq{2cm}{0cm}{r = 27.5 \pm 0.4.}

As for the remainders, we use an estimate based on general arguments about the convergence 
of the chiral series \cite{DescotesGenon:2003cg}
		
\myeq{1cm}{0cm}{\label{Delta_G}\Delta_G^{(4)}\ \approx\pm 0.3 G,\quad \Delta_G^{(6)}\ \approx\pm 0.1 G,}

\no where $G$ stands for any of our 2-point or 4-point Green functions,
which generate the remainders. We implement (\ref{Delta_G}) by using a normal distribution with $\mu\,$=\,0 and $\sigma\,$=\,0.3$G$ or $\sigma\,$=\,0.1$G$ for the NLO or NNLO remainders, respectively. The remainders are thus limited only statistically, not by any upper bound.

At last, we are left with three free parameters: $X$, $Z$ and $R$. These
control the scenario of spontaneous breaking of chiral symmetry and isospin breaking in our results. In the case of $X$ and $Z$, we use the constraint from the so-called paramagnetic inequality \cite{DescotesGenon:1999uh} and assume these parameters to be in the range

\myeq{1.25cm}{0cm}{0 < X < X(2),\quad 0< Z < Z(2).}

\no For the two flavor order parameters, we use the lattice QCD values listed in table \ref{tab1} \cite{Bernard:2012fw}. In addition, we implement a constraint following from $X(2),Z(2)>$\,0, similarly to \cite{DescotesGenon:2003cg}.

We use two approaches to deal with $R$. In the first one we assume it to be a known quantity. We use the value

\myeq{2cm}{0cm}{R=37.8\pm3.3,}

\no obtained from a dispersive analysis of $\eta$$\,\to\,$$3\pi$ \cite{Kampf:2011wr}. However, one should be aware that this estimate is based on an assumption that NNLO standard $\chi$PT \cite{Bijnens:2007pr} converges well at a specific kinematic point found in unphysical region. In comparison, \cite{Bijnens:2007pr} arrives at $R$\,=\,42.2, Dashen's theorem at leading order gives $R$\,$\sim$\,44, while at NNLO $R$\,$\sim$\,37. Alternatively, we leave $R$ free, or more precisely, assume it to be in a wide range $R$\,$\in$\,(20,\,60).

We resort to Monte Carlo sampling in order to perform the numerical integration in (\ref{Bayes}). We have used 10000 samples per grid element, the total number of samples being $\sim\,$$4\cdot10^6$.

\begin{figure}[t]
	\hsp{0cm}
	\includegraphics[width=0.52\textwidth]{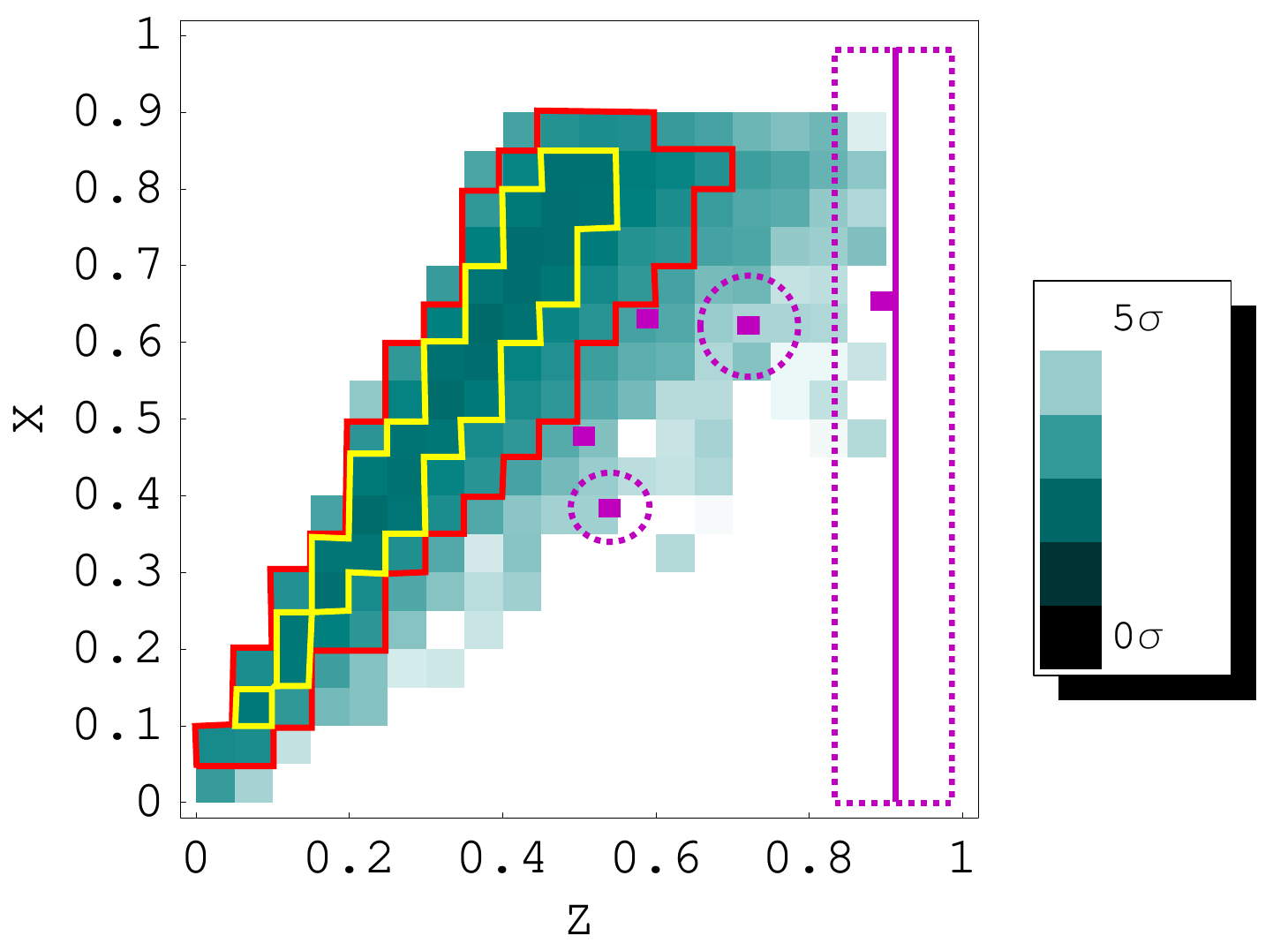}
	\caption{Probability density $P(X,Z|\mathrm{data})$ for $R$\,=\,37.8$\,\pm\,$3.3\newline
					 highlighted areas: yellow: 1$\sigma$ C.L. contour, red: 2$\sigma$ C.L. contour \newline
					 \hsp{2.1cm}purple: results listed in table \ref{tab2}}
	\label{fig1}
\end{figure}

\begin{figure}[t]
	\hsp{0cm}
	\includegraphics[width=0.5\textwidth]{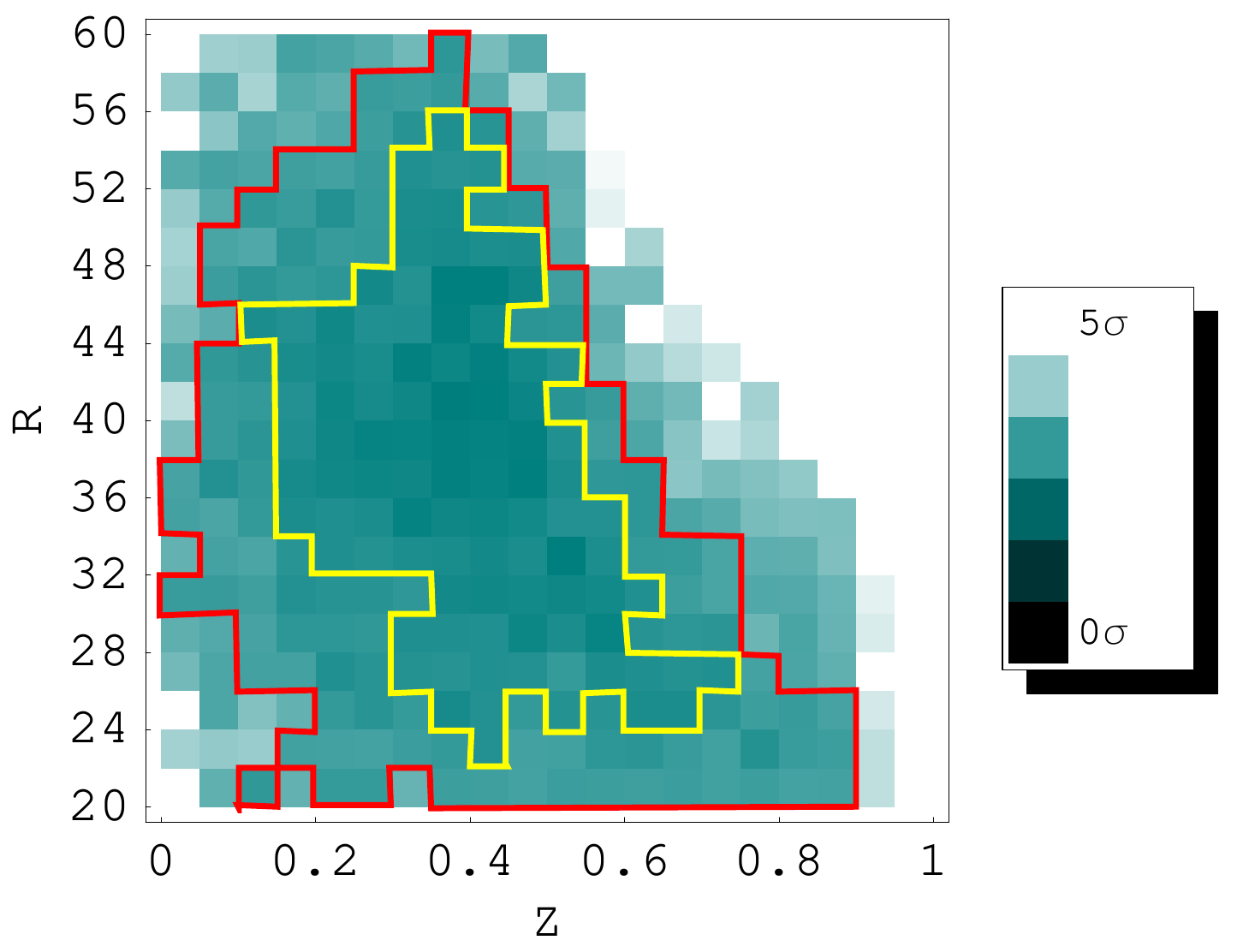}	
	\caption{Probability density $P(R,Z|\mathrm{data})$ for $R$ free, $X$ integrated out\newline
					 highlighted areas: yellow: 1$\sigma$ C.L. contour, red: 2$\sigma$ C.L. contour }
	\label{fig2}
\end{figure}

The obtained probability density distributions can be found in figures \ref{fig1} and \ref{fig2}. As can be seen, when assuming $R$\,=\,37.8\,$\pm$\,3.3, there is some tension with available results. The $\eta$$\,\to\,$$3\pi$ data seem to prefer a larger ratio of chiral order parameters $Y$\,=\,$X/Z$\,$\sim$\,1.5 than recent $\chi$PT and lattice fits and appear to rule out large values of $Z$.

As expected, it's hard to constrain $R$ without information on $X$ and $Z$. Even in this case a chunk of the parameter space can be excluded at 2$\sigma$ C.L., e.g. simultaneously moderate or large values of $R$ and $Z$. When integrating $R$ out, we obtain $0.05 < Z <0.75$ at 2$\sigma$ C.L.

\vsp{-0.2cm}																						





\nocite{*}
\bibliographystyle{elsarticle-num}
\bibliography{Bibliography}








\end{document}